\documentclass{article} 
\usepackage{iclr2021_conference,times}

\usepackage{amsmath,amsfonts,bm}

\def\eqref#1{equation~\ref{#1}}

\def\1{\bm{1}}

\DeclareMathAlphabet{\mathsfit}{\encodingdefault}{\sfdefault}{m}{sl}
\SetMathAlphabet{\mathsfit}{bold}{\encodingdefault}{\sfdefault}{bx}{n}

\usepackage{hyperref}
\usepackage{url}

\usepackage{graphicx}
\usepackage{subfigure}

\title{Prediction of Tuberculosis using U-Net and segmentation techniques}

\author{Dennis Núñez-Fernández, Gabriel Jiménez-Avalos, \\ 
\textbf{Jorge Coronel, Patricia Sheen, Mirko Zimic} \\
Laboratorio de Bioinformatica y Biología Molecular \\
Universidad Peruana Cayetano Heredia, Peru \\
\texttt{\{dennis.nunez,gabriel.jimenez,jorge.coronel} \\
\texttt{patricia.sheen,mirko.zimic\}@upch.pe} \\
\And
Lamberto Ballan \\
Visual Intelligence and Machine Perception Group \\
University of Padova, Italy \\
\texttt{\{lamberto.ballan\}@unipd.it} \\
}

\iclrfinalcopy 
\begin{document}

\maketitle

\begin{abstract}
One of the most serious public health problems in Peru and worldwide is Tuberculosis (TB), which is produced by a bacterium known as Mycobacterium tuberculosis. The purpose of this work is to facilitate and automate the diagnosis of tuberculosis using the MODS method and using lens-free microscopy, as it is easier to calibrate and easier to use by untrained personnel compared to lens microscopy. Therefore, we employed a U-Net network on our collected data set to perform automatic segmentation of cord shape bacterial accumulation and then predict tuberculosis. Our results show promising evidence for automatic segmentation of TB cords, and thus good accuracy for TB prediction.
\end{abstract}

\section{Introduction}

Worldwide, around 1.7 billion people, equivalent to a quarter of the world’s population, are infected with Mycobacterium tuberculosis; and 5–10\% of them will develop tuberculosis (TB) along their lifetime. Also, it is estimated that 1.2 million people died from tuberculosis in 2018 \citep{1_}. The MODS method (Microscopic Observation Drug Susceptibility Assay) \citep{2_} allows the growth and recognition of morphological patterns of mycobacteria in a liquid medium directly from a sputum sample. It is a fast and low-cost alternative test, since in only 7 to 21 days TB can be detected. This method has been added to the list of rapid tests authorized by the National Tuberculosis Prevention and Control Strategy. In addition, using lens-less microscopy contributes to its automation because it is easier to use and calibrate by no trained personnel.

In recent years a deep learning approach, the Convolutional Neural Networks (CNNs), have become the state-of-the-art for object recognition in computer vision \citep{3_}, and have high potential in image classification \citep{3_} and object detection \citep{4_,5_}. In the field of medical image analysis using computer vision techniques, the automatic segmentation has been taken an important role. Manual segmentation by experienced clinicians is very important and priceless; however, it is laborious and time-consuming, and may be subjective. In the last years, an approach for image segmentation, a CNN model known as U-Net, has shown promising results \citep{6_} in medical image analysis. The U-Net has been applied in microscopy images, including brain tissue characterization and segmentation \citep{10_}, vessel wall segmentation \citep{8_}, detection and counting of cells \citep{7_}, and identification of bacteria \citep{11_}.

The purpose of this work is to evaluate the feasibility of fully automatic segmentation of TB cords in lens-free images by applying the U-Net. Therefore, with the implementation of this project, we will have a faster and easier diagnostic tool for TB, which would be used by more people.

\clearpage

\section{Methodology}

\begin{figure}[h]
\vskip 0.2in
\begin{center}
\centerline{\includegraphics[width=0.94\columnwidth]{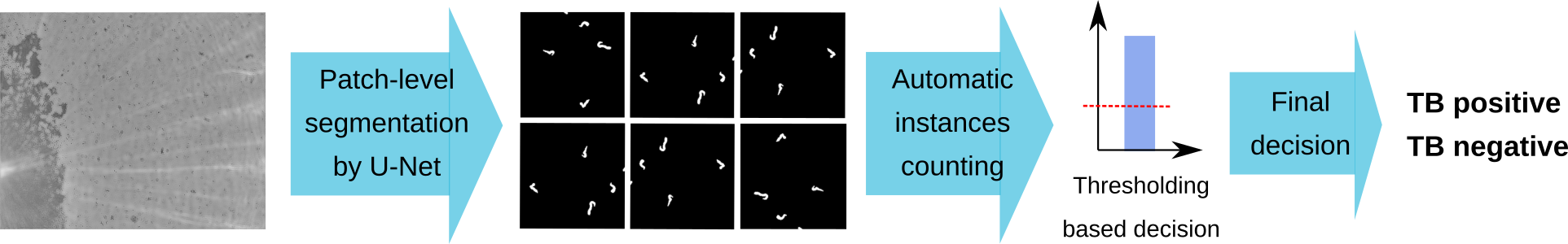}}
\caption{Overview of the methodology.}
\label{overview}
\end{center}
\vskip -0.2in
\end{figure}

Our proposed methodology is based on making use of an automatic segmenting of Tuberculosis (TB) cords using a convolutional neural network in order to have a more robust and efficient automatic segmenting device compared to traditional segmentation techniques. Therefore, the proposed methodology has the following steps: First the total image of 3840x2700 pixels is divided into sub-images of 256x256 pixels, then each sub-image is processed through the proposed U-Net architecture, which gives the automatic segmentation of TB cords in each sub-image. Later, these sub-images are concatenated to reconstruct the automatic segmentation of TB cords for the full image. Finally, an automatic counting of the TB cords is carried out, and based on a predefined threshold value, the final decision is made. The steps described above are shown in Figure~\ref{overview}.

The images were collected by the iPRASENSE Cytonote lens-free microscope. In this way, the images are obtained based on the reconstruction from the holograms that were obtained by the microscope. The holograms are captured by a CMOS sensor located as close as possible to the sample. Finally, at data collection, we obtain 10 full images with a dimension of 3840x2700 pixels each one and presented in grayscale format.

\begin{figure}[h]
\vskip 0.2in
\begin{center}
\centerline{\includegraphics[width=0.29\columnwidth]{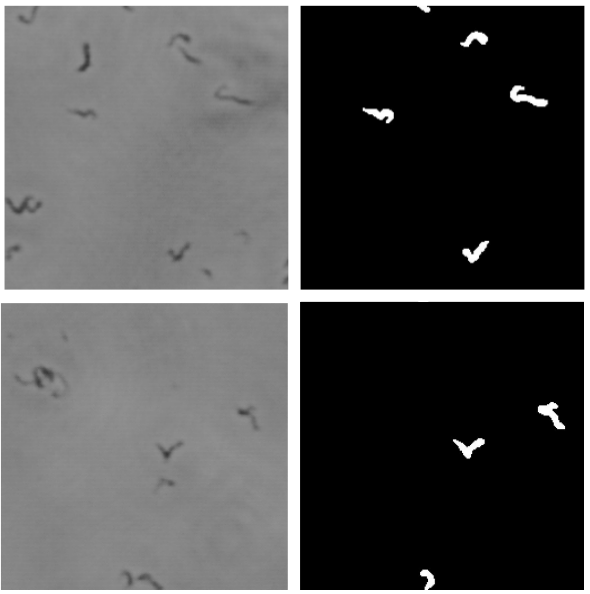}}
\caption{Samples of the generated dataset.}
\label{samples}
\end{center}
\vskip -0.2in
\end{figure}

For the generation of the dataset, the following steps were carried out: First a medical expert (with extensive experience in TB research) highlighted the TB cords in a rectangle. Then, a staff, which was trained to recognize the morphology of TB cords, delineated the edges of the TB cords. After, the sub-images and their TB cord masks were obtained using simple computer vision techniques. The generated dataset has 150 grayscale and 150 binary sub-images of 512x512 pixels size, similar to the samples depicted in Figure~\ref{samples}.

\begin{figure}[h]
\vskip 0.2in
\begin{center}
\centerline{\includegraphics[width=0.90\columnwidth]{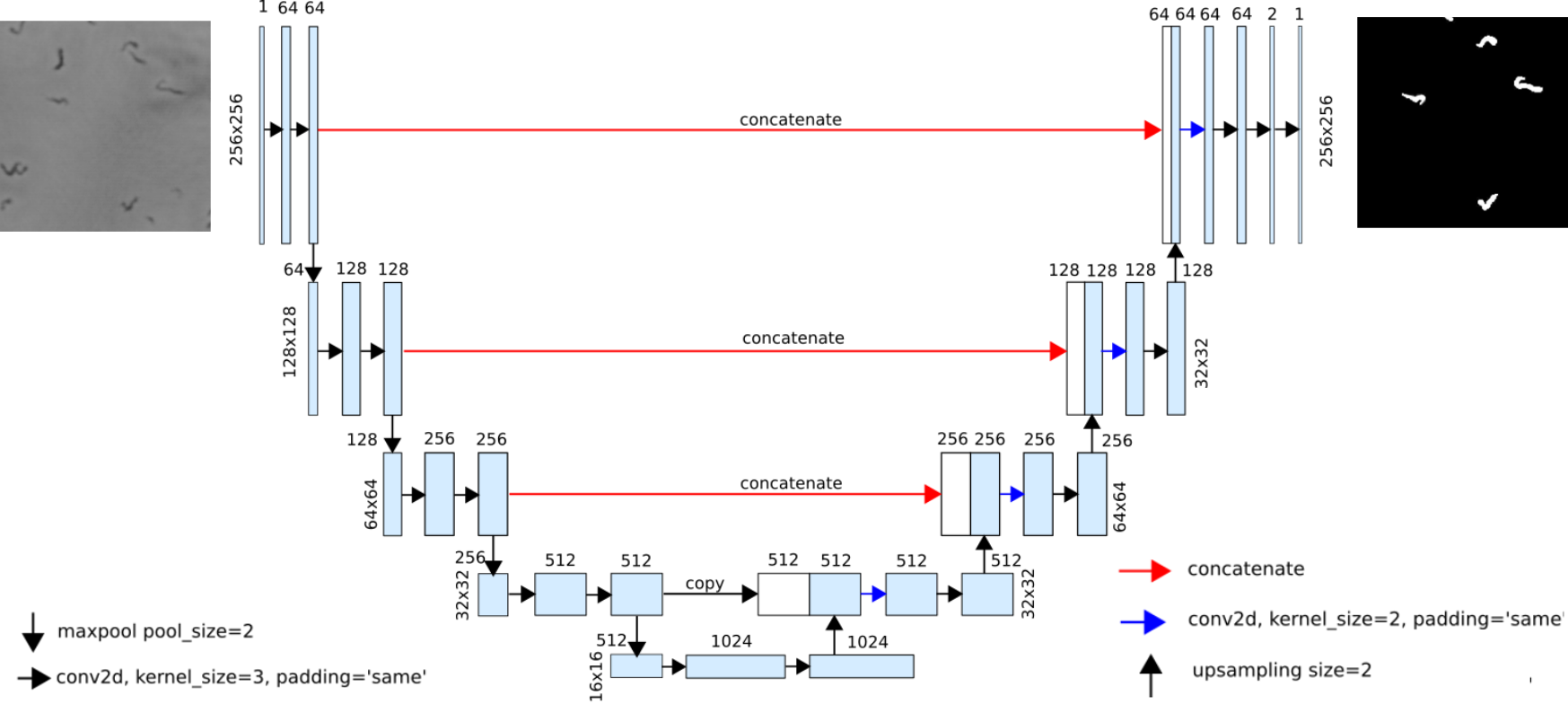}}
\caption{The U-Net architecture used for segmentation.}
\label{unet}
\end{center}
\vskip -0.2in
\end{figure}

We make use of the U-Net architecture and some hyper parameters were modified in order to work with our collected dataset and to produce an accurate segmentation. As Figure~\ref{unet} shows, the U-Net consists of a contract and an expansive path. The contraction section is made of many contraction blocks. Each block takes an input applies two 3X3 convolution layers followed by a 2X2 max pooling. During the contraction, the spatial information is reduced while feature information is increased. The expansive path combines the feature and spatial information.  Each block passes the input to two 3X3 CNN layers followed by a 2X2 upsampling layer. At the final layer a 1x1 convolution is used to map each 64-component feature vector to the desired number of classes. In total the network has 23 convolutional layers.

Once the regions of TB cords are obtained, dilation and erosion operations are performed to connect the nearest regions and have one region per object. Then automatic counting of TB cords is carried out using the connected components algorithm. Finally, based on a predefined threshold value, the final decision is made. The predefined threshold value is obtained based on the decision of the TB expert by examining all the images. Thus, TB patients have a number of cords much higher than the threshold value, and patients without TB have a much lower number of cords. We obtained a threshold value of 10 instances, this means that if the automatic counter gives a value greater than 10, it is a case of positive TB, otherwise it is a case of negative TB.

\section{Results}

For the evaluation of the U-Net architecture, we used the intersection-over-union (IoU or Jaccard Index) metric since it provides a better performance measure compared with pixel accuracy. We employed 120 sub-images for training the and 30 sub-images for testing. 

\begin{figure}[h]
\vskip 0.2in
\begin{center}
\centerline{\includegraphics[width=0.54\columnwidth]{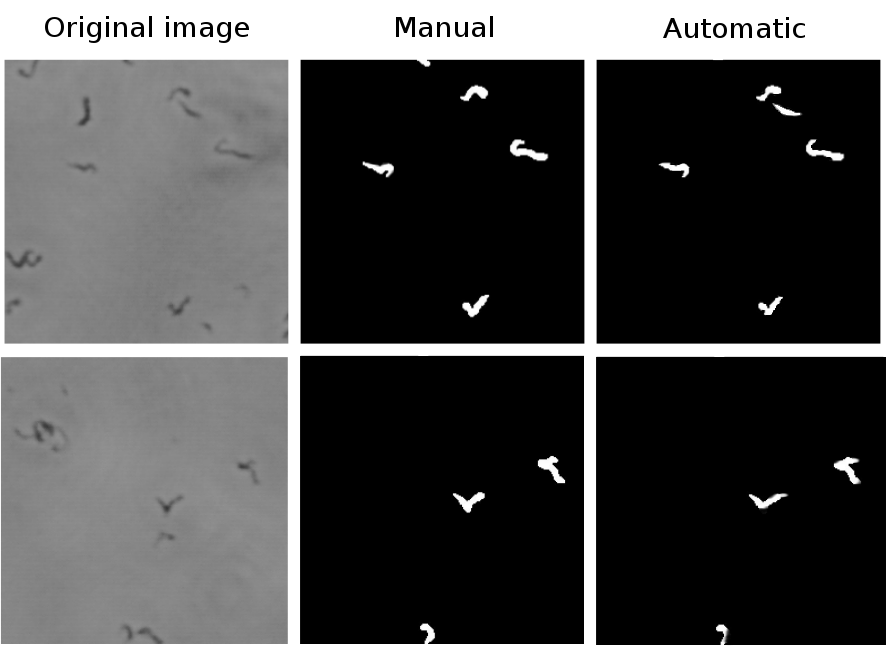}}
\caption{Results of the automatic segmentation for the U-Net.}
\label{results}
\end{center}
\vskip -0.2in
\end{figure}

For the U-Net architecture, we obtain an IoU of 0.88. The results show a good performance despite the fact that the dataset is quite noisy and has a resolution that makes it difficult to define the TB cords. As you can see in Figure~\ref{results}, the automatic segmentation looks quite similar to the manual segmentation. Some regions of the automatic segmentation are wrong segmented due to the low resolution and the difficulty to define TB cords in such areas.

In addition to the previous experiment, we compared the performance of the U-Net architecture with another techniques such as the K-Means clustering \citep{16__}, Full Connected Network (FCN) for segmentation \citep{12__}, SegNet architecture \citep{14__}, and the UNET++ architecture \citep{15__}. See Table~\ref{cnns-table}. As can be seen, and as expected, the U-Net architecture achieves a better performance compared to classical techniques such as K-Means or FCN, but a similar performance to U-Net++ or Segnet, which are based on convolutional networks. In the same way, we verified that the total performance of the system to predict TB cases, achieves an accuracy of 86.7\% for the U-Net case, higher than the other methods performed.

\begin{table}[h]
\caption{Performance for different approaches and CNN architectures}
\label{cnns-table}
\begin{center}
\begin{tabular}{lll}
\multicolumn{1}{c}{\bf APPROACH}  & \multicolumn{1}{c}{\bf IOU ACCURACY (\%)} & \multicolumn{1}{c}{\bf TOTAL ACCURACY (\%)}
\\ \hline \\
K-Means \citep{16__}  &  42.2$\pm$ 4.6  &  40.2 $\pm$ 7.5 \\
FCN \citep{12__}  &  71.2 $\pm$ 5.1  &  65.2 $\pm$ 2.2 \\
SegNet \citep{14__}  &  82.7 $\pm$ 3.8  &  79.2 $\pm$ 6.8 \\
UNET++ \citep{15__}  &  86.5 $\pm$ 4.2  &  84.2 $\pm$ 2.3 \\
\textbf{U-Net \citep{6_}}  &  \textbf{88.4 $\pm$ 3.6}  &  \textbf{86.7 $\pm$ 4.8} \\
\end{tabular}
\end{center}
\end{table}

\section{Conclusions}

In this work we demonstrated that our system is capable for automatic segmentation of tuberculosis cords with an Intersection over Union (IoU) of 0.88 and a total accuracy of 86.7\%. This by using images captured from a lens-free microscopy and using the U-Net architecture. These experiments show promising results despite the fact that the project is still under development, i.e. more samples are still being collected and some improvements are still being implemented. In addition, the use of segmentation makes the system explainable, as it shows the regions for which the system has made such a decision; thus in case there is a low difference with the decision threshold value, the physician or specialist can quickly verify. The most important fact is that this project contributes to a fast, cost-effective, and globally useful tool for tuberculosis detection, especially in the Peruvian rural areas where medical resources are limited.

\clearpage

\bibliography{references}
\bibliographystyle{iclr2021_conference}

\end{document}